\documentstyle[epsfig]{mn}
\begin{document}

\title
[The anomalous properties of Markarian 1460]  
{The anomalous properties of Markarian 1460} 

\author[Neil Trentham, R.~Brent Tully and Marc A.~W.~Verheijen ]  
{
Neil Trentham$^{1}$, R.~Brent Tully$^{2}$ and 
Marc A.~W.~Verheijen$^{3,4}$ \\ 
$^1$ Institute of Astronomy, Madingley Road, Cambridge, CB3 0HA.\\ 
$^2$ Institute for Astronomy, University of Hawaii,
2680 Woodlawn Drive, Honolulu HI 96822, U.~S.~A.\\ 
$^3$ NRAO-Array Operations Center, P.~O.~Box 0, Socorro NM 87801, U.~S.~A.\\
$^4$ Raman Research Institute, C.V. Raman Ave., Sadashivanagar, 
Bangalore 560 080, India 
}
\maketitle 

\begin{abstract} 
{ 
We present and discuss
optical, near-infrared and HI measurements of the galaxy
Markarian 1460 at a distance of
19 Mpc in the Ursa Major Cluster.  This low-luminosity ($M_B = -14$)  
galaxy is unusual because 
(i) it is blue ($B-R=0.8$) and has the spectrum of an HII galaxy, 
(ii) it has a light profile that is smooth and well fit by an $r^{1/4}$
and not an exponential function at all radii larger than the seeing, and 
(iii) it has an observed central brightness of about $\mu_B = 20$ mag
arcsec$^{-2}$, intermediate between those of elliptical galaxies 
(on the bright $\mu_B$ side) and
normal low-luminosity dwarf irregular 
(on the low $\mu_B$ side) galaxies.  
No other known galaxy exhibits all these properties in conjunction.
On morphological grounds
this galaxy looks like a normal
distant luminous elliptical galaxy, since
the fundamental plane tells us that
higher luminosity normal elliptical galaxies tend to have
lower surface-brightnesses.
Markarian 1460 has $2 \times 10^7$ M$_{\odot}$ of HI
and a ratio M(HI)/L$_B$ of 0.2,
which is low compared to typical values for star-forming dwarf galaxies.
From the high surface brightness and $r^{1/4}$ 
profile, we infer that the baryonic
component of Markarian
1460 has become self-gravitating through dissipative
processes.  From the colours, radio continuum,
HI and optical emission line properties, yet smooth texture,
we infer that Markarian 1460 has had significant star formation as recently as
$\sim 1$ Gyr ago but not today.
}
\end{abstract} 


\section{Introduction} 

The global photometric properties of normal galaxies can be summarized by
the following parameters:  
(1) absolute magnitude, (2) colours, (3) compactness, and (4) 
radial distribution and scale-length.
To a first approximation, high-luminosity giant
galaxies are either early-type (Elliptical or SO)
and have a radial profile that is well-fit by a de Vaucouleurs (1948) 
$r^{1/4}$ law or
late-type (spiral or irregular) 
and have a radial profile that is well-fit by an exponential law
(Freeman 1970). 
Lower luminosity galaxies (dwarf irregulars and dwarf spheroidals) also
have exponential light profiles 
(Binggeli \& Cameron 1991) but have lower surface brightnesses. 
Many late-type galaxies have light profiles which can be decomposed into
a bulge ($r^{1/4}$) and disk (exponential) part (Kormendy 1977, Kent 1985). 

More subtle effects can be seen by examining correlations between the
various parameters for a sample of galaxies (see Fig.~1):
\vskip 1pt \noindent
(1) Early type galaxies are redder and have older stars than late-type ones.
\vskip 1pt \noindent
(2) Luminous early-type galaxies tend to have higher 
surface-brightnesses than luminous late-type
galaxies.  They are more compact (there do
exist, however, late-type galaxies with compact cores);
\vskip 1pt \noindent
(3) Late-type galaxies tend to have lower surface-brightnesses as their
luminosity decreases.  This trend continues into the regime of
dwarf galaxies;  
\vskip 1pt \noindent
(4) There is a tendency for more luminous early-type galaxies to have lower
surface-brightnesses.  This is characterized by the familiar fundamental
plane for elliptical galaxies (Kormendy \& Djorgovski 1989); 
\vskip 1pt \noindent
(5) Most low-luminosity galaxies tend to be late-type galaxies in the field
and dwarf spheroidal galaxies in clusters; 

\begin{figure*}
\begin{minipage}{170mm}
{\vskip-3.5cm}
\begin{center}
\epsfig{file=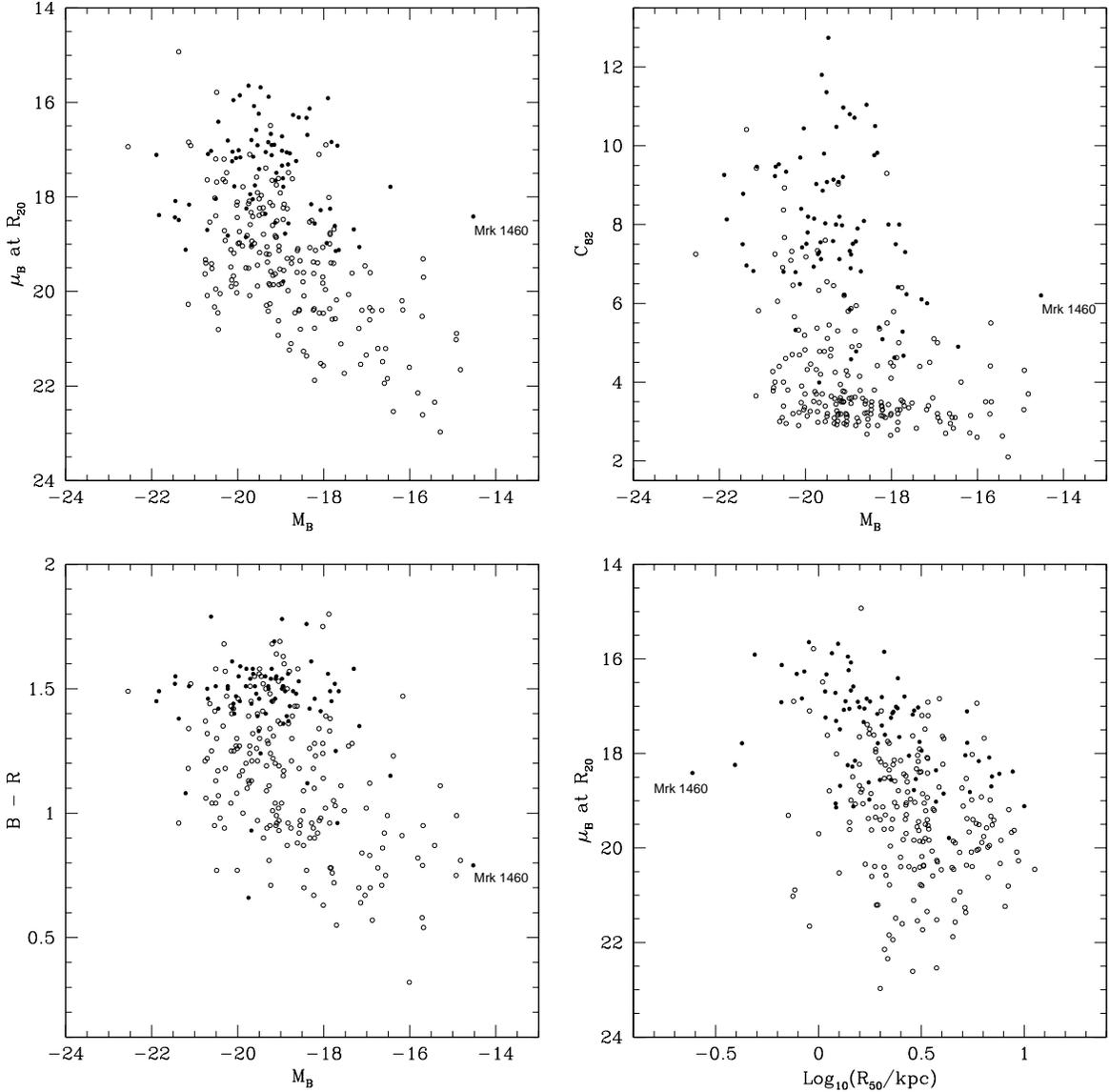, width=18.65cm}
\end{center}
{\vskip-5.7cm}
\caption{Photometric parameter correlations for early-type (filled
circles; Hubble Type $T \leq 0$) and late-type (open circles; $T > 0$) 
galaxies from the samples of
Pierce \& Tully (2001; local field, Virgo and Fornax galaxies)
and Tully et al.~(1996; Ursa Major galaxies). 
The panels are: upper-left -- absolute $B$ magnitude $M_B$ 
{\it vs.~}effective surface brightness $\mu_B$ within an a radius
$R_{20}$ containing
20\% of the light of the galaxy;
upper-right -- $M_B$
{\it vs.~}concentration index $C_{82}$, the ratio of the radii containing
80\% and 20\% of the total light; 
lower-left -- $M_B$ {\it vs.~}$B-R$ colour;
lower-right -- half-light radii $R_{50}$ {\it vs.~} $\mu_B$. 
Mrk 1460 is indicated in each panel and is classified as an elliptical
galaxy by virtue of its $r^{1/4}$ profile (see the text) although its
colours are more characteristic of late-type galaxies.
} 
\end{minipage}
\end{figure*}

\noindent
There do exist other, rarer galaxies which are not well-described by the
parameterizations used for normal galaxies. 
Examples are huge low surface-brightness giants like those 
studied by Sprayberry et al.~(1995), peculiar interacting galaxy systems
like Arp 220 (Sanders \& Mirabel 1996), and blue  
compact dwarfs, 
low-luminosity galaxies which appear to have been caught in a short burst of
extreme star formation.

We now present observations of another kind of galaxy that appears to be
anomalous.  It is Markarian 1460
(hereafter Mrk 1460), which
is a low-luminosity
($M_B = -14.5$)
blue emission-line galaxy with a $r^{1/4}$ light
profile.
This galaxy is anomalous since it looks morphologically like an elliptical
galaxy and has no obvious sign of clumpiness (as do most blue emission-line
low-luminosity galaxies which also have exponential light profiles;
Telles, Melnick \& Terlevich 1997).
Its observed central surface-brightness is also much higher than that of
most low-luminosity galaxies, about 
$\mu_R$ = 20 mag arcsec$^{-2}$.
This surface-brightness is
characteristic of the most luminous elliptical galaxies.  
Mrk 1460 is, however, much bluer than normal ellipticals.
Mrk 1460 was identified in a spectroscopic
survey of galaxies with ultraviolet
continuum by Markarian et al.~(1984), 
who showed that it was an HII galaxy in the
Ursa Major Cluster of galaxies 
(distance modulus = 31.35; Tully \& Pierce 2000).
It was
subsequently included in the
photometric survey of the Ursa Major Cluster of Tully et al.~(1996).
Its compactness was noticed by Tully \& Verheijen (1997) who classify
it as a blue example of a
compact dwarf (``Type 5" in their notation).  

In Section 2 of this paper we
describe the observations of this galaxy from our recent optical and HI
surveys, and present the photometric properties in Section 3 and discuss
them in the context of the radio and spectroscopic measurements.  In
Sections 4 through 6 we discuss similarities between Markarian 1460
and various other astronomical objects
and possible formation scenarios.

\section{Observations and data reduction}

\subsection{Optical} 

\begin{figure}
\begin{center}
\vskip 1mm
\psfig{file=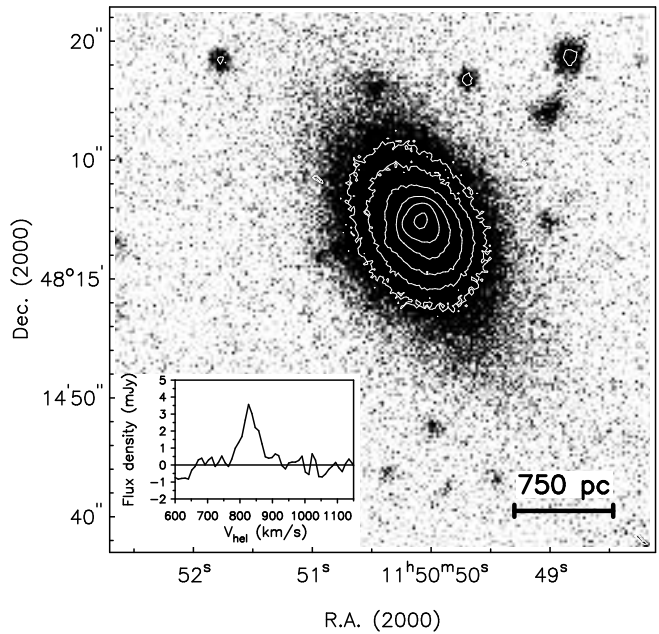, width=8.65cm}
\end{center}
\vskip -3mm
\caption{
$R$-band image of
Markarian 1460 from the data of Trentham et al.~(2001).
The insert is the HI spectrum from the data of Verheijen et al.~(2001).}
\end{figure}

In this work we use optical $BRI$-band images of Mrk 1460
from Tully et al.~(1996)
and deeper $R$-band data from Trentham, Tully \& 
Verheijen (2001).  The relevant
telescopes and
instruments, observing configurations, data-reduction strategies,
and details of the photometric systems
are given in those papers.
The total magnitudes computed by Tully et al.~(1996) are:
$B=16.89$, $R=16.06$ and $I$=15.75.
These values have been updated to include Galactic extinction corrections
from the data of Schlegel, Finkbeiner \& Davis (1998) and do not include
a correction for internal extinction. 
An $R$-band image of this galaxy is presented in Figure 2.

Old corrected apparent magnitudes  :   16.83 16.04 15.73   (Table 5)

Of course, in the end it doesn't matter at all but you may want to
drop the internal extinction correction and use the new Schlegel et al
correction for Galactic extinction:

New Galactic extinction (Schlegel) :    0.09  0.06  0.04
New corrected apparent magnitudes  :   16.89 16.06 15.75

\subsection{Near Infrared}

Mrk 1460 was imaged through a $K^{\prime}$
filter (Wainscoat \& Cowie 1992) using the QUIRC
1024 $\times$ 1024 InSb array (Hodapp et al.~1996,
scale 0.19$^{\prime \prime}$ pixel$^{-1}$,
field of view $3.2^{\prime} \times 3.2^{\prime}$)
at the 2.24 m University of Hawaii Telescope
on Mauna Kea on the night of January 23, 2000.
Conditions were photometric for these observations with a median seeing
of about 0.7$^{\prime \prime}$ FWHM.
Exposures were as a sequence of five, one-minute frames, dithered
in order to
reject bad pixels.  Offset fields were observed with
equal exposure time to the
target fields in order to perform a sky subtraction.
Flat-field images were constructed using
dome flats and sky images were constructed
from the offset blank sky frames.
Individual frames were flat-fielded, sky-subtracted, and then registered
and combined.
Instrumental magnitudes were computed from observations of
5 $-$ 10 UKIRT faint standards (Casali \& Hawarden 1992),
giving a photometric accuracy of about 2\%.
A $2 \sigma$ isophotal magnitude 
from this image was measured as $K^{\prime}
=14.92$.  This will be somewhat fainter than the total magnitude since
we are missing flux at large radii.  If we estimate this flux from
extrapolating the best-fit exponential light profile (as was done by 
Tully et al.~1996), the flux at
large radius accounts for 0.58 magnitudes and the total 
magnitude is $K^{\prime}
=14.34$.  For the best-fitting $r^{1/4}$ light profile (see Section 3.1 and 
Figure 3 below), the flux at large radius accounts for 0.87 magnitudes and
the total magnitude is $K^{\prime}
=14.05$.

\subsection{Radio}

Mrk 1460 was observed with the VLA in its D-configuration by
Verheijen et al.~(2000, 2001) as part of a blind HI survey of the Ursa
Major Cluster.  The details of the observations are given there.  The
galaxy was detected in HI (see Fig.2 for the HI spectrum) and the
observed width of the HI profile is 92 km s$^{-1}$ at the 20\% level.
The total HI mass is $2 \times 10^7$ M$_\odot$ and the 1.4 GHz
continuum flux density is 0.77 $\pm$ 0.43 mJy.

The emission is spatially unresolved given the 45$^{\prime\prime}$
beam and it is unclear whether the width of the HI profile can be
related to ordered rotation of a gas disk of unknown inclination in
the gravitational potential of the galaxy.  Consequently, estimating
the dynamical mass from the width of the HI profile is not feasible.
However, suppose the HI gas is distributed in a rotationally
supported disk at an inclination of 60 degrees and that the maximum
rotational velocity is reached at 1 kpc from the center which is
roughly the observed optical extent.  Correcting the width of the
global HI profile for instrumental and turbulent broadening and
inclination gives a rotational velocity of 41 km/s which translates to
a total dynamical mass of $4 \times 10^8$ M$_\odot$ within a 1 kpc
radius.

\section{Observed Properties of Markarian 1460}

\subsection{Light distribution} 

\begin{figure}
\begin{center}
\vskip-2mm
\epsfig{file=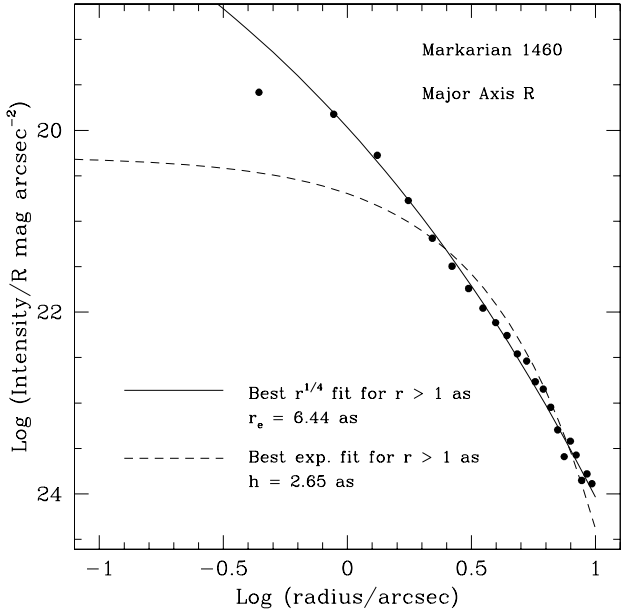, width=8.65cm}
\end{center}
\vskip-5mm
\caption{
The
major-axis $R$-band light profile 
of Mrk 1460, from the data of 
Trentham et al.~(2001).
The filled circles represent the data.  
The solid line represents the
best-fitting $r^{1/4}$ profile to the profile at
radii greater than one arcsecond (the seeing FWHM).  The
dashed line represents the best-fitting exponential profile over
the same radius range. 
}
\end{figure}

As mentioned in the previous section, Mrk 1460 appears on a CCD image
similar to
a distant elliptical galaxy by virtue of its $r^{1/4}$ light profile
and its central surface brightness of about 19.5 $R$ mag arcsec$^{-2}$, which
is characteristic of the central surface-brightnesses of luminous
elliptical galaxies.
It is far more compact than any of the other dwarf irregular
galaxies in the Ursa Major Cluster samples of Trentham et al.~(2001) and
Verheijen et al.~(2001).

The light profile is well fit by an $r^{1/4}$ profile (Fig.~3) in all
passbands.  It is smooth and only deviates from elliptical symmetry in the  
central 1.5 arcseconds; the flux peaks slightly to the northeast of the center
as defined by the outer isophotes. 
Additionally, there is a modest and very low surface-brightness 
excess of light to the southwest. 
There is no sign of 
substantial clumpiness as is seen in the other low-luminosity galaxies
in the Ursa Major sample of Trentham et al.~(2001) with HI detections.
An exponential law fails to fit the galaxy simultaneously
at both small and large radii, at a high level of significance:
the reduced $\chi^2$ ($\nu$ = 10 degrees of freedom) for the $r^{1/4}$ and
exponential fits in Figure 3 are 0.85 and 131 respectively. 
The central surface-brightness derived from the
$r^{1/4}$ fit at
radii larger than one arcsecond (corresponding to the seeing radius) is  
$\mu_{0R}$ = 15.2 mag arsec$^{-2}$.
This extrapolated value is very much higher than
the {\it measured} central surface brightnesses of
about $\mu_{0R}$ = 19.5 mag arsec$^{-2}$.

\subsection{Current star-formation rate and metallicity}

We can infer the star formation rate in Mrk 1460 from measurements
of the OII line equivalent width
(46 \AA; Pustilnik et al.~1999) and also the 1.4 GHz 
continuum flux (Section 2.3, Verheijen et al.~2001).  
Both sets of measurements directly probe high-mass stars and their remnants
and derivations of the star formation rate from these measurements therefore
require very substantial corrections for low-mass stars, which
dominate the total mass.  
Assuming a Salpeter (1955) stellar initial mass function and the
OII line calibration of Gallego (1998), the current star formation rate
in Markarian 1460 is 
$0.09 \,{\rm M}_{\odot} {\rm yr}^{-1}$. 
Assuming the same stellar initial mass function and the
calibration of Cram et al.~(1998), the 3$\sigma$ upper limit on the
star formation rate derived from the 1.4 GHz measurements described in
Section 2.3 is $0.11 \,{\rm M}_{\odot} {\rm yr}^{-1}$.  

Deriving abundances from emission-line properties is complicated and a
detailed analysis requires many more measurements than are available
for this galaxy (Stasinska \& Leitherer 1996).
However, the following simple analysis is suggested.
From the flux ratios O[II]/H$\beta$ and
O[III]/H$\beta$ (Pustilnik et al.~1999),
the abundance ratios O$^{+}$/H and O$^{++}$/H  are
3.7 $\times$ 10$^{-5}$ and 4.6 $\times$ 10$^{-5}$,
leading to a
value of the ionized gas oxygen abundance of O/H = 8.3 $\times$ 10$^{-5}$.
In this calculation we follow
Tully et al.~(1981) and adopt a normal Whitford (1958) reddening curve.
If the heavy element abundance is proportional to the
oxygen abundance, the total metallicity of the ionized gas in
Mrk 1460 is then about 0.1 solar (Anders \& Grevesse 1989).

\subsection{Colours}

The colours of this galaxy are: $B-R=0.83$, $R-I=0.31$,
and
$I-K^{\prime}$=1.91.  The $I-K^{\prime}$ colour is derived using
aperture magnitudes within the $K^{\prime}$ 2-$\sigma$ isophote; this
ensures that we are measuring the same part of the galaxy in both filters
and does not require us to make corrections for light lost below the sky
at large radius in the $K^{\prime}$ image.

These optical colours suggest an age of approximately 1.3 Gyr
if the galaxy has been forming stars either in an instantaneous burst at this
time in the past or continuously with an exponential
star-formation history profile
with $e$-folding time 1 Gyr, given the models
of Bruzual \& Charlot (1993), assuming a
Salpeter (1955) stellar initial mass function from 0.1 M$_{\odot}$
to 100 M$_{\odot}$),
negligible internal extinction
and a metallicity of 0.4 solar.
The $I-K^{\prime}$ 
colour above is, however, too red by about 0.7 magnitudes to be
produced by the stars from this burst alone.  This might suggest the
presence of a population of older stars which only contribute in a small
way to the optical fluxes.  Probably the detailed star-formation history
of this galaxy is too complex to derive from broadband colours alone, but
it appears that a single recent star formation episode is unlikely since 
the optical--near infrared
colour is so red.  Indeed, the $B-K^{\prime} = 3.1$ colour of
this galaxy is redder than that of
31 of the 33 galaxies with Hubble Type $T \geq 6$
in the Ursa Major sample of Tully et al.~(1996) that have near-infrared
photometry, all of which have blue optical colours.  
 
The colour gradients that we measure in this galaxy are fairly small:
${\rm d} (B-R) / {\rm d} r \approx
0.27\, {\rm mag\,arcsec}^{-1}$
between 1 and 2 arcsec and
${\rm d} (B-R) / {\rm d} r \approx
0.13\, {\rm mag\,arcsec}^{-1}$
between 2 and 4 arcsec
(note that 12 arcsec corresponds to 1 kpc at the distance of this
galaxy) along the major axis.
The central colour that we measure within the seeing FWHM radius
is $B-R \sim 0.5$, about 0.3 magnitudes bluer than for the galaxy
as a whole.  This might suggest that the stars in the very center of
the galaxy are slightly younger than average stars in the galaxy (we will
return to this point in Section 5). 

\subsection{Total baryonic content}

The total
$B$-band solar luminosity is $L_B = 1.0 \times 10^8 {\rm L}_{\odot B}$,
and the HI gas mass is
$M({\rm {HI}}) = 2 \times 10^7 \,{\rm M}_{\odot}$.
The ratio $M({\rm {HI}}) / L_B$ is then 0.2, which is lower than the
values normally seen for very late-type emission-line galaxies
(Young \& Knezek 1989, Verheijen \& Sancisi 2001).  
It therefore appears that Mrk 1460 is deficient
in HI for a normal star-forming dwarf galaxy.

Given the star-formation rate and age in the previous two sections,
the total stellar mass of Mrk 1460 is $7 \times 10^7 \,{\rm M}_{\odot}$.
This suggests that the HI gas only contributes
about 20 per cent of the total baryonic mass.
This last number depends on both the shape of the
stellar initial mass function at low masses and on the normalization of
Gallego's (1998) calibration, both of which are very uncertain.
However the low value of $M({\rm {HI}}) / L_B$
does suggest that the baryonic mass of
Mrk 1460 is not heavily dominated by cold atomic gas.

The total baryonic mass is then about $7 \times 10^7 \,{\rm M}_{\odot}$
(the mass in stars) plus $2 \times 10^7 \,{\rm M}_{\odot}$ (the mass in
HI) plus $7 \times 10^7 \,{\rm M}_{\odot}$ (accounting for the helium
mass fraction in the cold gas), totalling 
$1 \times 10^8 \,{\rm M}_{\odot}$.  This is comparable to the dynamical
mass estimated in Section 2.3, which suggests that the baryons 
are self-gravitating, at least in the center of the galaxy.   

\section{Discussion}

In this section we discuss similarities and
differences between Markarian 1460 and
other low-luminosity galaxies.  We then argue that galaxies like this
are rare and suggest possible reasons for why it is so different to the
majority of low-luminosity galaxies.

Low luminosity galaxies can be discussed in the context of their surface
brightness and age properties.  
\vskip 1pt \noindent
(i) Tully \& Verheijen (1997) have suggested
that the separate regimes of low and high surface brightness can be 
distinguished on the basis of dynamical evidence as follows.  In low 
surface-brightness systems the observed baryonic component cannot account for the
observed rotation with reasonable mass-to-light choices and it is inferred 
that dark matter is dynamically dominant even near the galaxy centers.
In high surface brightness systems the observed light and a reasonable
association of mass-to-light comfortably accounts for the inner galaxy rotation
and it can be concluded that the baryonic component is self-gravitating within
the inner couple of exponential scalelengths; 
\vskip 1pt \noindent
(ii) Systems with young
populations still have cold gas and are still forming stars
or have only just recently completed a major episode of star
formation.  They are blue, gas-rich
and tend to have clumpy light distributions.   
Systems with only old populations do not have a gas reservoir and the
old stars have little or no memory of the precise location of their birthplaces due
to stellar-dynamical effects like phase mixing and violent relaxation.
 
Most low-luminosity systems with young populations
are dwarf irregular galaxies, which are low
surface-brightness systems.  
A detailed study of the morphologies of low-luminosity star-forming
galaxies is performed by
Telles et al.~(1997).  They found that the majority of
their sample galaxies had azimuthally-averaged
exponential light profiles (see their Section 3.5). 
This was also true of all the other
galaxies in the Ursa Major sample of Trentham et al.~(2001).  Most of the
star-forming galaxies in
both the Telles et al.~sample and in the Ursa Major sample also
showed evidence for lumpy, irregular morphologies.  

Low suface-brightness galaxies with only old stars
are dwarf spheroidal galaxies, which
are red, gas-poor, and have smooth light profiles that are also
exponential.  This is the type that is so
numerous at faint magnitudes in the Virgo (Phillipps et al.~1998)
and Fornax (Kambas et al.~2000) Clusters.  The similarity in  
profile shapes and surface brightnesses (Binggeli \& Cameron 1991, Binggeli
1994) between dwarf irregular and spheroidal 
galaxies is suggestive of an evolutionary link: for example,
dwarf spheroidals may be dwarf irregulars that
have blown out any residual gas via supernova-driven winds at some time
in the distant past (e.g.~Kormendy \& Bender 1994).  

Red compact dwarfs are also gas-poor and consist 
of old stars, but these have high surface-brightnesses.  
These normally have $r^{1/4}$ light profiles.  The scatter in central
surface-brightness of such objects is large, ranging from about 13 $B$
mag arcsec$^{-2}$ for M32 (Binggeli \& Cameron 1991) to about 19  
$B$ mag arcsec$^{-2}$ for
UGC 6805 in the Ursa Major Cluster (Tully et al.~1996). 

Markarian 1460 is the 
rare example of a high surface-brightness galaxy caught in
transition {\it between young and old}.
Like red compact dwarfs, it has
a smooth $r^{1/4}$ light profile.  Its central surface-brightness of 
20.7 $B$ mag arcsec$^{-2}$ is at the faint end of the range of
central surface-brightnesses for elliptical galaxies and red compact
dwarfs, hence its appearance similar to that of a distant luminous
elliptical galaxy (consider just the filled circles
in the upper left panel of Fig.~1). 
Like dwarf irregular galaxies, Mrk 1460 has young
stars and an emission-line spectrum (Pustilnik et al.~1999).  
Blue compact dwarf (BCD)
galaxies with $r^{1/4}$ light profiles have been known
to exist for some time (e.g.~Kunth, Maurogodarto \&
Vigroux 1988).
Doublier (1998) places a number of these objects on a
magnitude vs.~central surface-brightness plot, and finds that they lie
close to the highest surface-brightness red compact dwarfs like M32,
several mag arcsec$^{-2}$ brighter than Mrk 1460. 
A well-studied example is NGC 1510
(Disney \& Pottasch 1977, Kinman 1978), which also shows
strong emission lines. 
Unlike Mrk 1460, however, this galaxy exhibits considerable
clumpiness (Eichendorf \& Nieto 1984). 
More generally, BCDs possess localized regions undergoing intense
star formation burst where the surface brightnesses
is extremely high.  
In Mrk 1460, however, the young stars exist on a
galactic scale and are not just 
restricted to any particular star-forming
region.

Galaxies with the properties of Mrk 1460 have not turned up in
field optical spectroscopic (e.g.~Ellis
et al.~1996, Lin et al.~1996) or HI (e.g.~Zwaan 1998)  
surveys (however see Drinkwater et al.~1988 for some possible
counterparts in the Fornax Cluster).  They are presumably very rare.
This is not surprising, since the current star-formation rate of
0.1 M$_{\odot}$ yr$^{-1}$ exhausts the HI gas in $2 \times 10^8$ yr,
which is small compared to a Hubble time.
It is likely that Mrk 1460 will evolve into an old high
surface-brightness dwarf, albeit one at least two magnitudes fainter
than UGC 6805 once fading is taken into account
(this would be the lowest luminosity $r^{1/4}$ galaxy known).  
Its final surface brightness will probably be close to what it is
now: Mrk 1460 does not possess enough cold gas
(unless it is in molecular form, but there is no
$IRAS$ detection at this position) to convert to stars via an
extreme dissipative
collapse, which is what would be required were Mrk 1460 to turn into
a very high surface-brightness red compact dwarf like M32.

So what happened to cause this galaxy to have these anomalous properties?
Making $r^{1\over{4}}$ galaxies by
violent relaxation on timescales short compared to stellar population
evolutionary timescales does not seem to be a problem (Lynden-Bell 1967,
van Albada 1982).
However, why this happened in this particular case, and not in most 
low-luminosity star-forming galaxies (which have exponential profiles and
lower surface brightnesses) is unclear. 
It might be an environmental effect: 
Tully \& Verheijen (1997) suggest
that tidal interactions 
can radially scramble the baryonic mass, presumably through gas collisions.
The baryonic matter transferred inward during
an encounter with another galaxy can become sufficiently concentrated that
it becomes self-gravitating.  A self-gravitating disk will then further
rearrange itself into a radially stable configuration (Mestel 1963).  The
result is a high concentration of baryons
towards the center of
the galaxy, which in turn leads to the high surface brightness relative to what
is seen in most low-luminosity galaxies.  
A related possibility is that Mrk 1460 formed out of a dense, self-gravitating
gas cloud assembled via a hydrodynamic shock process (Barnes \& Hernquist
1992), perhaps in proximity to another galaxy in the Ursa Major Cluster.

More generally, understanding the physical mechanisms at work
would be helped significantly by obtaining a sample of objects
like Mrk 1460.  Such galaxies are distinctive in spectroscopic
surveys since their
surface brightnesses are high and their emission lines are strong.
This will shortly be possible as large samples of
galaxies with known redshifts and morphological information
become available from the new generation
of wide-field deep redshift surveys like the Sloan Digital Sky
Survey and the 2DF Survey.


\begin{thebibliography}{}

\bibitem[\protect\citename{bl}%
]{and}
Anders E., Grevesse N., 1989, Geochim.~Cosmochim.~Acta, 53, 197

\bibitem[\protect\citename{bl}%
]{bir}
Barnes J.~E., Hernquist L., 1992, Nat, 360, 715

\bibitem[\protect\citename{bl}%
]{bir}
Binggeli B., 1994, in Meylan G., Prugneil P., ed., ESO
Conference and Workshop Proceedings No.~49: Dwarf Galaxies. 
European Space Observatory, Munich, p.~13

\bibitem[\protect\citename{bl}%
]{bi2}
Binggeli B., Cameron L.~M., 1991, A\&A, 252, 27

\bibitem[\protect\citename{bl}%
]{br}
Bruzual G., Charlot S., 1993, ApJ, 405, 538

\bibitem[\protect\citename{bl}%
]{ca}
Casali M.~M, Hawarden T.~G., 1992, The JCMT-UKIRT Newsletter Vol.~4, p.~33

\bibitem[\protect\citename{bl}%
]{cr}
Cram  L., Hopkins A., Mobasher B., Rowan-Robinson M., 1998, ApJ, 507, 155

\bibitem[\protect\citename{bl}%
]{de}
de Vaucouleurs G., 1948, Ann.~d'Ap., 11, 247

\bibitem[\protect\citename{bl}%
]{di}
Disney M.~J., Pottasch S.~R., 1977, A\&A, 60, 43

\bibitem[\protect\citename{bl}%
]{do}
Doublier V., 1998, in Thuan T.X., Balkowski C., Cayatte V.,
    Tran Thanh Van J., eds, Dwarf Galaxies and Cosmology.  Editions
    Fronti{\`{e}}res, Paris, p.~163

\bibitem[\protect\citename{bl}%
]{dr}
Drinkwater M.~J., Gregg M.~D., Smith R.~M.,
1998, in Thuan T.X., Balkowski C., Cayatte V.,
    Tran Thanh Van J., eds, Dwarf Galaxies and Cosmology.  Editions
    Fronti{\`{e}}res, Paris, p.~163

\bibitem[\protect\citename{bl}%
]{ei}
Eichendorf W., Nieto J.-L., 1984, A\&A, 132, 342

\bibitem[\protect\citename{bl}%
]{el}
Ellis R.~S., Colless M., Broadhurst T., Heyl J., Glazebrook K.,
1996, MNRAS, 280, 235

\bibitem[\protect\citename{bl}%
]{fr}
Freeman K.~C., 1970, ApJ, 160, 811

\bibitem[\protect\citename{bl}%
]{ga}
Gallego J., 1998, in Thuan T.X., Balkowski C., Cayatte V.,
    Tran Thanh Van J., eds, Dwarf Galaxies and Cosmology.  Editions
    Fronti{\`{e}}res, Paris, p.~31


\bibitem[\protect\citename{bl}%
]{ho}
Hodapp K.~W.~et al.~1996, New Astron., 1, 77

\bibitem[\protect\citename{bl}%
]{kam}
Kambas A., Davies J.~I., Smith R.~M., Bianchi S., Haynes J.~A., 2000, 
AJ, 120, 1316  

\bibitem[\protect\citename{bl}%
]{ke}
Kent S.~M., 1985, ApJS, 59, 115 


\bibitem[\protect\citename{bl}%
]{kin}
Kinman T.~D., 1978, AJ, 83, 764

\bibitem[\protect\citename{bl}%
]{ko}
Kormendy J., 1977, ApJ, 217, 406 

\bibitem[\protect\citename{bl}%
]{kb}
Kormendy J., Bender R., 1994, in Meylan G., Prugneil P., ed., ESO
Conference and Workshop Proceedings No.~49: Dwarf Galaxies.
European Space Observatory, Munich, p.~161

\bibitem[\protect\citename{bl}%
]{kd}
Kormendy J., Djorgovski S., 1989, ARA\&A, 27, 235 

\bibitem[\protect\citename{bl}%
]{ku}
Kunth D., Maurogodarto S., Vigroux L., 1988, A\&A, 204, 10

\bibitem[\protect\citename{bl}%
]{li}
Lin H., Kirshner R.~P., Shectman S.~A., Landy S.~D., Oemler A.,
Tucker D.~L., Schechter P.~L., 1996, ApJ, 464, 60

\bibitem[\protect\citename{bl}%
]{ly}
Lynden-Bell D., 1967, MNRAS, 136, 101

\bibitem[\protect\citename{bl}%
]{ma}
Markarian E., Lipovetsky V.~A., Stepanian J.~A., 1984, Astrofizica, 21, 419

\bibitem[\protect\citename{bl}%
]{me}
Mestel L., 1963, MNRAS, 126, 553

\bibitem[\protect\citename{bl}%
]{ph}
Phillipps S., Parker Q.~A., Schwartzenberg J.~M., Jones J.~B., 1998,
ApJ, 493, L59

\bibitem[\protect\citename{bl}%
]{pi}
Pierce M.~J., Tully R.~B., 2001, in preparation 

\bibitem[\protect\citename{bl}%
]{pu}
Pustilnik S., Engles D., Ugryumov V., Lipovetsky V.~A.,
Hagen H.$\,$J., Kniazev A.~Y., Izotov Y.~I., Richter G.,
1999, A\&AS, 137, 299

\bibitem[\protect\citename{bl}%
]{sa}
Salpeter E.~E., 1955, ApJ, 121, 161

\bibitem[\protect\citename{bl}%
]{sanmr}
Sanders D.\,B., Mirabel I.\,F., 1996, ARA\&A, 34, 749

\bibitem[\protect\citename{bl}%
]{sfd}
Schlegel D.~J., Finkbeiner D.~P., Davis M., 1998, ApJ, 500, 525 

\bibitem[\protect\citename{bl}%
]{sp} 
Sprayberry D., Impey C. D., Bothun G. D., Irwin M. J., 1995, ApJ, 109, 558

\bibitem[\protect\citename{bl}%
]{st}
Stasinska G., Leitherer C., 1996, ApJS, 107, 661

\bibitem[\protect\citename{bl}%
]{te}
Telles E., Melnick J., Terlevich R.~J.,
1997, MNRAS, 288, 78

\bibitem[\protect\citename{bl}%
]{tr}
Trentham N., Tully R.~B., Verheijen M.~A.~W., 2001, MNRAS, in press
(astro-ph/0103039)  

\bibitem[\protect\citename{bl}%
]{tu1}
Tully R.~B., Boesgaard A.~M., Dyck H.~M., Schempp W.~V., 1981, ApJ, 246, 38

\bibitem[\protect\citename{bl}%
]{tu10}
Tully R.~B., Pierce M.~J., 2000, ApJ, 533, 744 

\bibitem[\protect\citename{bl}%
]{tu2}
Tully R.~B., Verheijen M.~A.~W., 1997, ApJ, 484, 145

\bibitem[\protect\citename{bl}%
]{tu3}
Tully R.~B., Verheijen M.~A.~W., Pierce M.~J., Huang J.$\,$-S.,
Wainscoat R.~J., 1996, AJ, 112, 2471

\bibitem[\protect\citename{bl}%
]{va}
van Albada T.~S., 1982, MNRAS, 201, 939

\bibitem[\protect\citename{bl}%
]{ver3}
Verheijen M.~A.~W. and Sancisi R., 2001, A\&A, in press (astro-ph/0101404)

\bibitem[\protect\citename{bl}%
]{ver1}
Verheijen M.~A.~W., Trentham N., Tully R.~B., Zwaan M.~A., 2000,
in Kraan-Korteweg R.~C., Henning P.~A., Andernach, H., eds, ASP
Conference Series 218: Mapping the Hidden Universe: The Universe Behind 
the Milky Way, The Universe in
HI, 
ASP, San Francisco, p.~263 

\bibitem[\protect\citename{bl}%
]{ver2}
Verheijen M.~A.~W., Trentham N., Tully R.~B., Zwaan M.~A., 2001, in
preparation

\bibitem[\protect\citename{bl}%
]{wai}
Wainscoat R.~J., Cowie L.~L., 1992, AJ, 103, 332

\bibitem[\protect\citename{bl}%
]{whi}
Whitford A.~E., 1958, AJ, 63, 201

\bibitem[\protect\citename{bl}%
]{you}
Young J.~S., Knezek P.~M., 1989, ApJ, 347, L55

\bibitem[\protect\citename{bl}%
]{zwa}
Zwaan M., 1998, in Thuan T.~X., Balkowski C., Cayatte V.,
    Tran Thanh Van J., eds, Dwarf Galaxies and Cosmology.  Editions
    Fronti{\`{e}}res, Paris, p.~49

\end{thebibliography}
\end{document}